\documentclass[dvips,twocolumn,amssymb, amsmath, nofootinbib,tightenlines,nobibnotes, prl, 10pt]{revtex4}

\usepackage{epsf}            
\usepackage{graphicx}         
\usepackage{color}            
\usepackage{rotate}

\newcommand{\opL}{{\bf\it \hat{L}}}
\newcommand{\be}{\begin{equation}}
\newcommand{\ee}{\end{equation}}
\newcommand{\rev}[1]{(\ref{#1})}
\newcommand{\lab}[1] {\label{#1}}

\newcommand{\Umar}[1] {}
\newcommand{\PutFigure}[1]{\begin{center}
\begin{picture}(320,150)
\put(10,0){\rotate[r]{\resizebox{6cm}{!}{\includegraphics{#1}}}}
\end{picture}\\
\end{center}
\vspace*{-0.5cm}
}

\begin{document}
\definecolor{myblue}{rgb}{0,0,0.8}

\title{\bf Creation of Kink and Antikink Pairs Forced By 
Radiation}
\author{Tomasz Roma\'nczukiewicz\thanks{trom@th.if.uj.edu.pl}
       \\Institute of Physics,\\
       Jagiellonian University, Reymonta 4, Krakow, Poland}

\begin{abstract}
The interaction between kink and radiation in nonlinear 
one-dimensional real scalar field is investigated. The process of 
discrete vibrational mode excitation in $\phi^4$ model is considered.
The role of this oscillations in creation of kink and antikink is discussed. 
Numerical results are presented as well as some attempts of analytical 
explanations. An intriguing fractal structure in parameter space dividing regions with creation and without is also presented.
\end{abstract}
\maketitle
%\input{p1_1}
%%%%%%%%%%%%%%%%%%%%%%%%%%%%%%%%%%%%%%%%%%%%%%%%%%%%%%%%%%%%%%%%%%%%%%
%
\section{\bf{Introduction}}
%%%%%%%%%%%%%%%%%%%%%%%%%%%%%%%%%%%%%%%%%%%%%%%%%%%%%%%%%%%%%%%%%%%%%%
%

Topological defects are usually compact, static  solutions with finite energy of partial differential equations which properties are defined by field values at infinity. 
The simplest examples of topological defects are one-dimensional kinks ($\phi^4$ or sine-Gordon equation). 
Less trivial examples (vortices, domain walls, monopoles etc.) manifest themselves in various branches of physics \cite{Coslab1}, \cite{Coslab2} like particle physics \cite{QCD}, condensed matter physics, cosmology \cite{Cosmic}, \cite{Saka} and much more.

Because of their stability and localized energy density some topological defects have similar properties as particles (e.g. kinks in 1+1 dimensions or monopoles in 3+1 dimensions).
They can be created or annihilated as ordinary particles. 
They can also interact with each other, with radiation or external force.
\cite{Kiselev} brings a very nice example of how topological defects accelerate in a constant external force.

However, topological defects reveal some properties which make them very different from particles. 
In our previous paper \cite{trom} we have presented an unexpected behavior when the kink in $\phi^4$ was exposed to monochromatic radiation of scalar field coming from one direction. 
The nonlinear character of this interaction resulted in ``negative  radiation pressure'', the kink, for most of the frequencies of incoming wave, instead of being pushed by radiation, was in fact being pulled toward the source  of this radiation.
Similar feature reveals the second most often used example, the sine-Gordon soliton. 

Another peculiarity of topological defects occurred in a two kink collision process in $\phi^4$ theory \cite{Matzner} during which the kinks could be scattered back or annihilated. 
Let us stress out that similar process in sine-Gordon equation, because of its integrability is less interesting. 
If the SG solitons have enough energy they go through each other with nothing but a phase shift. 
When the energy is below some threshold the solution describe a bound state - a breather which is a perfectly periodic solution.
$\phi^4$ has much more interesting structure of solutions. 
During collision the kinks loose some of their initial energy on behalf of radiation and when the loss is small the kinks reflect from each other. 
When the loss is large they glue to each other forming a bound state (so called  oscillon \cite{Gleiser}). 
The bound state also radiates and finally vanishes.
An extraordinary thing about this process is that there exists no real threshold. 
The regions in phase space for one or the other scenario mix and form a fractal similar to a Cantor set.

This is not an isolated example. \cite{Weinstein} brings a different example when a sine-Gordon soliton interacts with a oscillating impurity mode. They also observed a fractal structure.

We found yet another example of fractal structure but in a quite different process. 
In $\phi^4$ the soliton possesses an internal degree of freedom - an oscillational mode (in \cite{Matzner} authors claim it is responsible for a peculiar behavior of kinks during collision). 
When that mode is excited it oscillates with a certain frequency. 
In linear approximation the mode oscillates infinitely long with constant amplitude. 
When one includes nonlinear correction one finds the mode couples to scattering spectrum and radiates reducing its amplitude and finally vanishes.
But when the initial amplitude is large enough so that its energy is just a little above a mass of two kinks, a kink and antikink are produced. 

In present work we investigated an opposite process. 
We light radiation far away from the kink from both directions so that the kink would stand still.
The radiation excites the oscillational mode and if the radiation amplitude and frequency is suitable the oscillational mode breaks up to kink and antykink.
Of course if radiation is large enough the kinks can be created even from vacuum, but because the oscillational mode gathers the energy our process seems to be much more efficient.
One surprising result we have found is that in a plane radiation amplitude vs radiation frequency the border separating solutions with and without creation is also a fractal. We claim that the nonlinear coupling between the vibrational mode and radiation is responsible for the structure.

Our paper is organized as follows. In the following section we give an introduction to the model discussed. We remind a spectral structure of linearization around the kink.
The next section is devoted to excitation of the oscillations mode. 
We present our numerical results and attempts of theoretical explanation. 
We present a simple model reproducing qualitative results of the full model.
Although on a precise level there are some discrepancies, the
general structure of solutions remains similar.
The last section is as usual conclusions and discussion.

We have encountered also a very interesting question of how one can
simplify and reduce a system from infinite number of degrees of
freedom to only  few.
\section{\bf The Model}

Let us consider one-dimensional real scalar field obeying the equation in
the form:
\be
  \ddot{\phi}-\phi''+U'(\phi)=0, \label{eqq1}
\ee
where $U(\phi)$ is a potential with at least two equal minima (vacua)
$\phi_v$ (without loss of generality we can assume that
$U(\phi_v)=0$) so that in these theories there exist static soliton
solutions $\phi_s$ described by the equation
\be
\phi_s(x):\;\;x-x_o=\pm\int_{\phi_{v_1}}^{\phi_{v_2}}\frac{d\phi}{\sqrt{2U(\phi)}}.
\ee
In this section we will discuss two the most known models $i.e.$
$\phi^4$ ($U(\phi)=\frac{1}{2}(\phi-1)^2$) and sine-Gordon
($U(\phi)=1-\cos(\phi)$) equations.
In these theories the static soliton solutions have the form:
\be
\left\{
\begin{array}{ll}
\phi_s(x)=\pm\tanh(x-x_o)&{{\rm for}\; \phi^4}\\
\phi_s(x)=\pm{\rm arc}\!\tan(e^{-(x-x_o)})& {\rm for\; S-G}
\end{array}\right.
\ee
Because of the translational invariance we can substitute $x_0=0$.
Let us add some small perturbation to the static kink solution
$(\phi=\phi_s+\xi)$. If the potential can be expanded in a Taylor' series (one can do so for many systems with an exception of certain compactons \cite{Arodz}).
\be
  U'(\phi)=U'(\phi_s+\xi)=U'(\phi_s)+
  U''(\phi_s)\xi+N(\xi,\phi_s)
\ee
where $N(\xi,\phi_s)$ is a nonlinear part in $\xi$.
We can write the equation for  $\xi$:
\be
  \ddot{\xi}-\xi''+V(x)\xi+N(\xi,x)\equiv\ddot{\xi}-\opL\xi+N(\xi,x)=0.
\lab{eq1}
\ee
where
\be
V(x)=U''(\phi_s(x))=
\left\{\begin{array}{ll}
{\displaystyle 1-\frac{2}{\cosh^2x}}& {\rm for\; S-G}\\
\\
{\displaystyle 4-\frac{6}{\cosh^2x}}&{\rm for\; }\phi^4\\
\end{array}\right..
\ee
We substitute $\xi(t,x)=e^{i\omega_kt}\eta_k(x)$ and find the
eigenvalues and the eigenfunctions of the operator
$\opL=-\frac{d^2}{dx^2}+V(x)$. We can divide the spectra into three
groups
\begin{enumerate}
\item translational zero modes $\eta_t=\phi_s'$: $\phi_s(x+\delta
x)\sim\phi_s(x)+\delta
x\phi_s'(x)$, $\omega_t=0$,
\item discrete oscillational modes (there is none for S-G and only one
for $\phi^4$ but in general there can be any finite number)
$$\eta_d^{(\phi^4)}(x)=\frac{\sinh x}{\cosh^2 x},\;\;\;\omega_d=\sqrt{3},$$
\item scattering modes:
$$
\eta_k(x)=
\left\{\begin{array}{ll}
e^{ikx}(ik-\tanh x)& {\rm for\; S-G}\\
e^{ikx}\left(3\tanh^2 x-1-k^2-3ik\tanh x\right)& {\rm for\;} \phi^4\\
\end{array}\right.,
$$
where $k^2=\omega^2-m^2$.
\end{enumerate}
It is very significant that scattering modes in these models have no 
reflection part. 
This is not a general feature. 
In fact it is quite rare, but for example spectra for $V(x)$ given in 
a form:
$$
V(x)=-\frac{N(N+1)}{\cosh^2x}
$$ are reflectionless for all integer $N$. There are exactly $N$ 
bounded modes. One of them is a translational mode and the rest of 
 them  are oscillational.

The existence of the oscillational mode in the $\phi^4$ theory is responsible for its nonintegrability. 
During for example collisions this mode is excited and takes some of the initial energy. 
Then the energy, due to the nonlinearities, is transfered to radiation modes.

Numerical results show that the radiation coming from infinity and hitting
the kink can cause the creation of the pair kink and antikink.
This process is much easier for $\phi^4$.
Even small amplitudes can force the process.
In sine-Gordon's case the process occurs for amplitudes compared to
the amplitudes when the creation can occur even far away form
kink in pure radiation.
The explanation is quite simple.
In $\phi^4$ there exist an oscillation mode which can be excited by
the radiation.
Manton \cite {Manton} investigated the creation of kink and antikink caused only by
a discrete mode.
When the energy contained in this mode was a little above then the
energy of two kinks a pair of kink and antikink was created and two kinks 
were radiated to infinity in opposite directions and antikink remained.
When the energy was not large enough the discrete mode oscillated with a
decreasing amplitude and the energy was radiated to infinity because
of the coupling to the scattering modes.
In the following section we will discuss an opposite process.
Radiation coming from infinity will couple to the oscillating mode.

\section{\bf Excitation of the oscillating mode and the creation process in $\phi^4$ model}
\subsection{Numerical results}

We used two different sets of initial and boundary conditions to 
determine the coupling between radiation and the oscillating mode.
In both cases we considered only antisymmetric problems, we evaluated 
our system only on one half of the $x$-axis and posed the boundary 
condition $\phi(t, 0)=0$.
We could do so because the symmetrical radiation does not couple with 
an oscillational mode and therefore gives no contribution. 
The symmetrical radiation couples however with a translational mode and 
the kink as a whole would start moving in one direction and  
would be a difficult object to study.

In the first case we had  initial conditions $\phi(0, x)=\phi_s(x)$.
The boundary conditions were $\dot{\phi}(0, x)=0$ and $\phi(t, L)=\frac{1}{2}A\sin\omega t$ (where $L$ was large in comparison to the
size of the kink, usually $L=200$). 
We use $\frac{1}{2}$ because when the waves come from both sides of the kink, they add and produce a standing wave with twice as big amplitude.
Because of the condition $\phi(t, 0) = 0$ and the symmetry of the
equation $\phi\rightarrow-\phi,x\rightarrow-x$ we actually studied the
system with boundary conditions
$\phi(t, \pm L)=\pm \frac{1}{2}A\sin\omega t$. 
These conditions describe the kink in radiation coming from both sides from very far away.
In order to measure the excitation of the discrete mode we calculated
the value (a projection on a discrete mode):
\be
A_d(t)=\frac{\displaystyle\int_{-\infty}^\infty\!\!\!dx\,\eta_d(x)\left[\phi(t, x)-\phi_s(x)\right]}{\displaystyle\int_{-\infty}^\infty\!\!\!dx\eta_d^2(x)}.
\label{Ad}
\ee

Because the radiation is orthogonal to $\eta_d$ there is no
contribution to $A_d(t)$ from any other modes (at least in linear 
approximation).
We measured $A_d$ for different $\omega$ and $A$.

\begin{figure}
\PutFigure{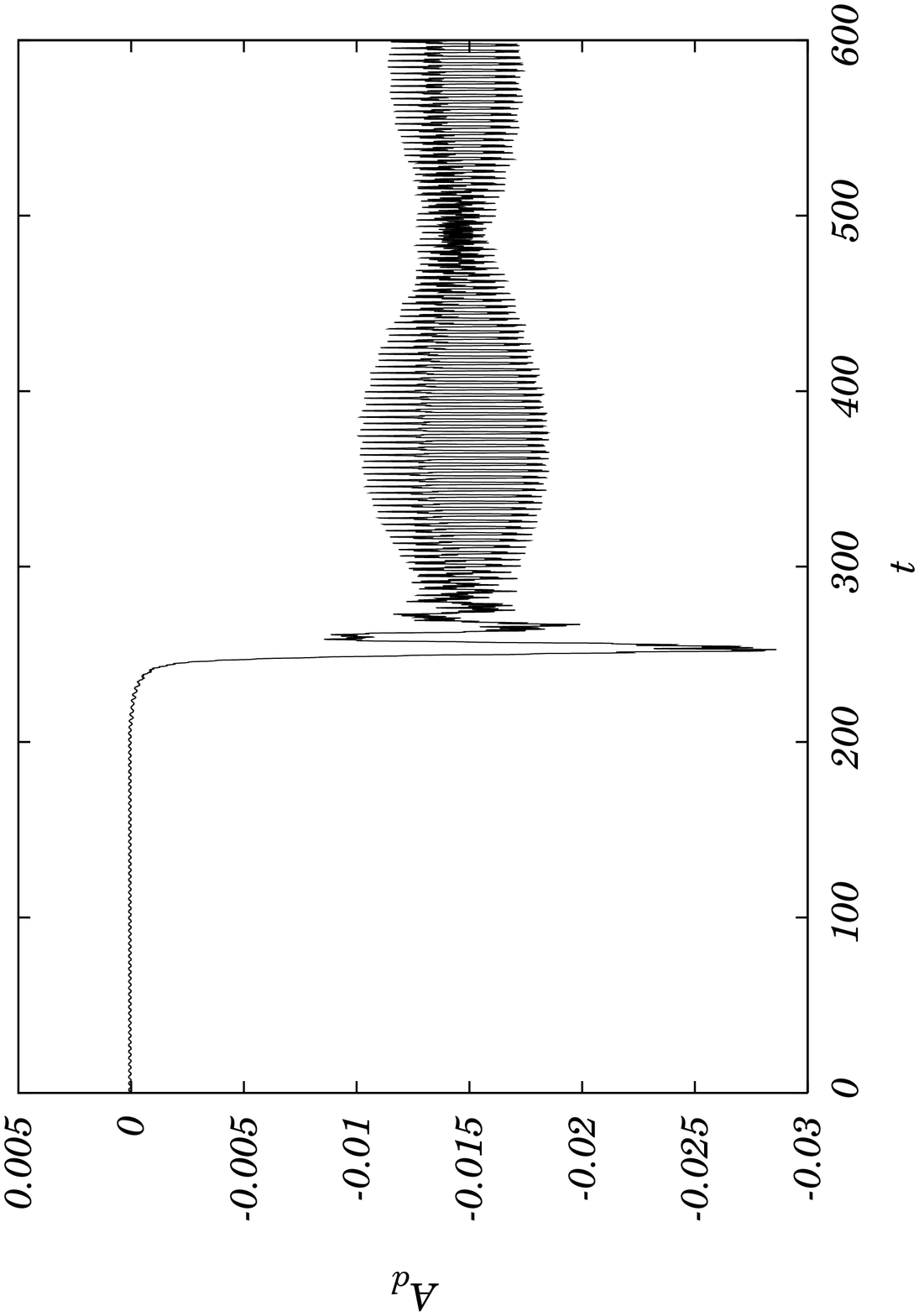}
\caption{The excitation of the discrete mode for $A=0.2$ and
$\omega=3.5$. \label{f1}
\vspace*{3mm}}
\end{figure}

\begin{figure}
\PutFigure{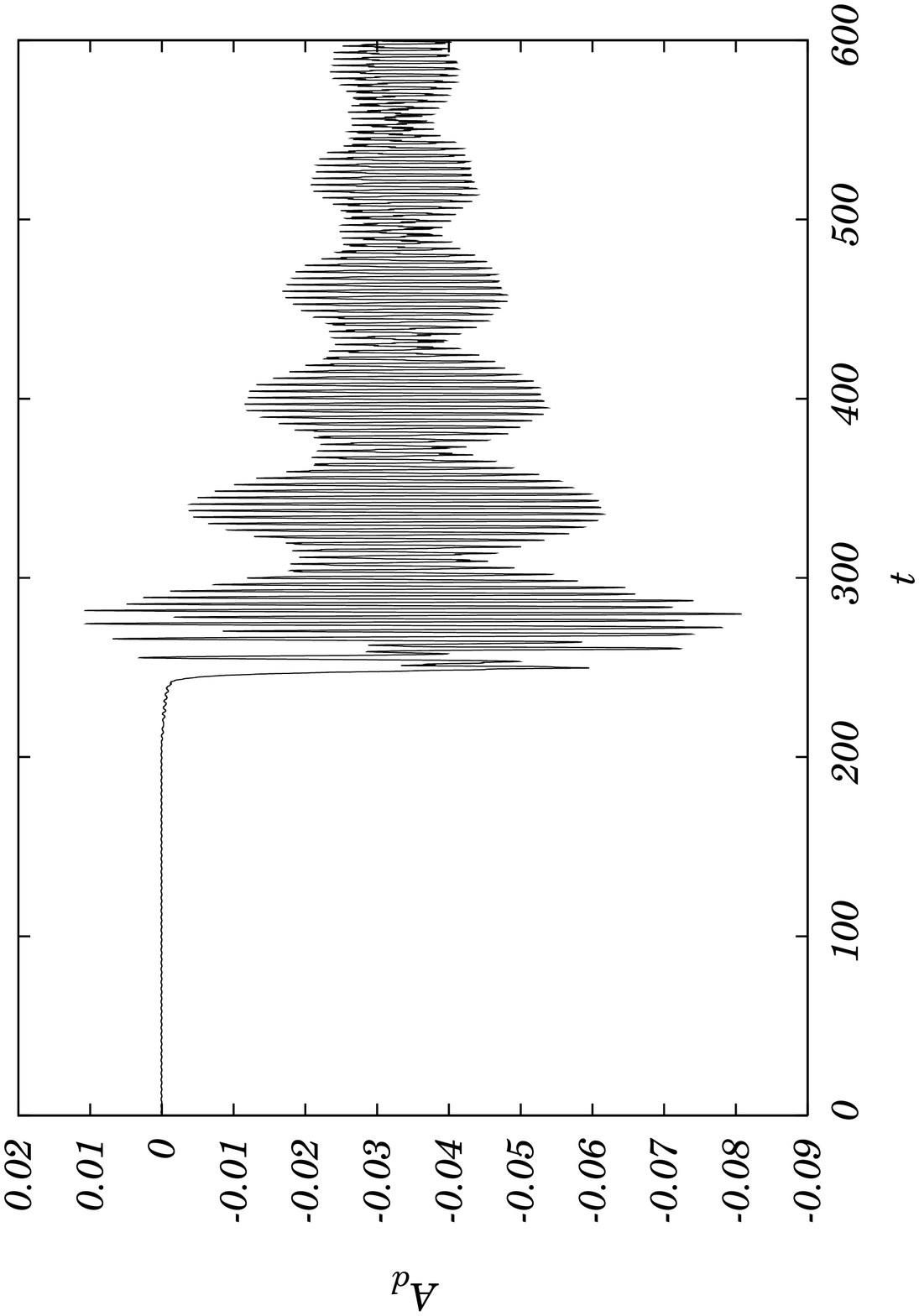}
\caption{The excitation of the discrete mode for $A=0.3$ and
$\omega=3.5$. \label{f2}
\vspace*{5mm}}
\end{figure}

\begin{figure}
\PutFigure{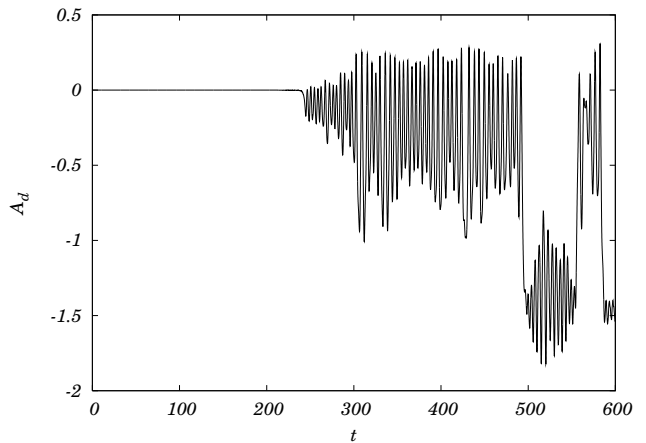}
\caption{The excitation of the discrete mode for $A=0.5$ and
$\omega=3.5$. We can see the creation of kink-antikink pairs for $t\approx500$, $t\approx560$ and $t\approx590$.
\label{f3}
\vspace*{5mm}}
\end{figure}

\begin{figure}
\PutFigure{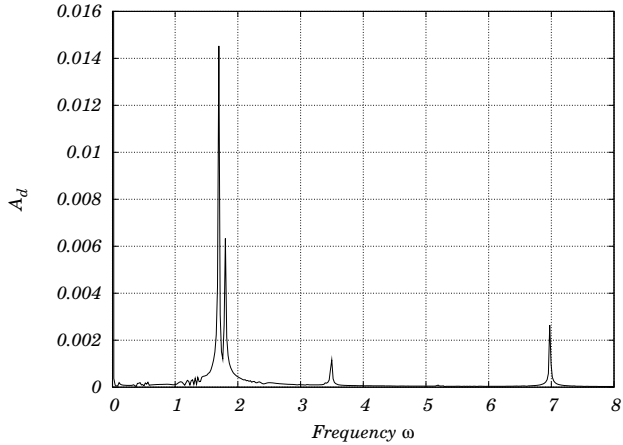}
\caption{Fourier transform of $A_d$ for $\omega=3.5$ and $A=0.2$.
One can see four separated peaks for
$\omega=\sqrt{3}\approx1.73,\;3.5,\;7$ and $3.5-\sqrt{3}\approx1.76$. \label{f4}
\vspace*{5mm}}
\end{figure}

\begin{figure}
%\PutFigure{P_040429_2}
\PutFigure{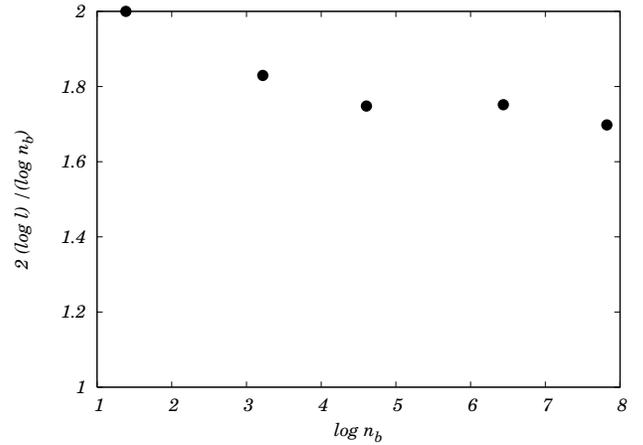}
\caption{Dependence $2\frac{\log l}{\log n_b}$ upon $\log n_b$ \label{f5_1}. We can expect that a fractal dimension would be more than 1.6.
\vspace*{5mm}}
\end{figure}

In figures \ref{f1}-\ref{f3} we have sketched three examples of $A_d$ $vs$ time for $\omega=3.5$ and for $A$ equals 0.2, 0.3 and 0.5.\\
For all Figures we can see that the reaction of the oscillating mode is retarded by about 240. Because the radiation is "switched on" at $x=L$ it needs time $t_0=\frac{\partial k}{\partial \omega}L=\frac{\omega L}{k}\approx243$ to get to the kink.\\
If the amplitude is small (Fig. \ref{f1}) $A_d$ oscillates with eigenfrequency
$\omega_d$. These oscillations are shifted and modulated by another frequency. As we will show in the following section this modulating frequency is $\omega_m=\omega-2\omega_d$ (Fig. \ref{f4}). In our example $\omega_m=0.0359$ and the period is $175$ so we can see a very good agreement.\\
If the amplitude is larger the modulations are with higher frequency, and the oscillations decay. This decay can be explained because as Manton showed in \cite{Manton} the oscillating mode couples to the scattering modes and radiates its energy (see also \cite{Pelka}, \cite{Slusarczyk}).\\
The next Figure \rev{f3} is the most interesting. First we can see some chaotic oscillations and then around $t=500$ $A_d$ suddenly goes to about $-1.5$ and then oscillates around that level, and then goes back to 0 (around $t=560$).
If the amplitude of $A_d$ is large enough a pair of kink and antikink can be created and radiated in both directions, but the place of a kink in the middle is taken by an antikink remains. When we substitute $\phi(x,t)=-\phi_s(x)$ in (\ref{Ad}) we  obtain that $A_d=-\frac{\pi}{2}\approx -1.57$.
\begin{figure}
%\PutFigure{P_040429_2}
\PutFigure{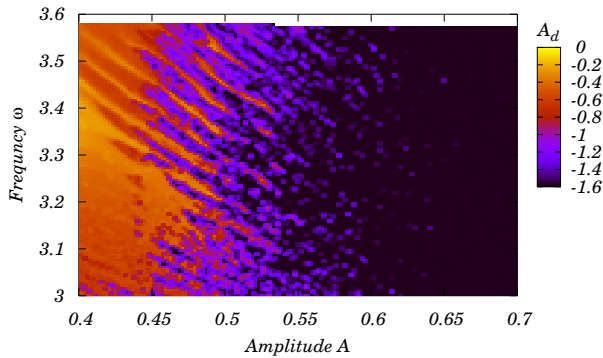}
\caption{Minima of $A_d$ vs frequency $\omega$ and amplitude $A$ of radiation coming from $L=200$ \label{f5}. Dark spots $A_d<-\frac{\pi}{2}\approx -1.57$ represent creation process.
\vspace*{3mm}}
\end{figure}
%In Figure \ref{f5} we have sketched the border between the solutions where pairs of kink and antikink are created and solutions where defects are not produced.
Figure \ref{f5} presents the minima of $A_d(t)$ as a dependence upon frequency $\omega$ and amplitude $A$ of incoming radiation. 
When $A_d$ is less than $-\frac{\pi}{2}$ a creation occurred (dark spots).
One can see the border between creation and small oscillations is very complicated, perhaps even fractal. 
The resolution is not good enough to prove this proposition but the Figure \ref{f5_1} may justify our hypothesis. 
We present the dependence 
\begin{equation}
  D(\log n_b) = 2\frac{\log l}{\log  n_b},
\end{equation}
where $n_b$ is a number of boxes and $l$ is a number of boxes containing the boundary. 
This value in limit $n_b\rightarrow\infty$ would give a true fractal dimension. 
Because we have only 2500 boxes we cannot tell the true dimension but it seems to be much larger than 1 somewhere around $1.6-1.7$ and that would justify that the border is fractal. 
Of course proving numerically that something is a fractal is very difficult and in our case is almost impossible.

\begin{figure}
%\PutFigure{P_040429_1}
\PutFigure{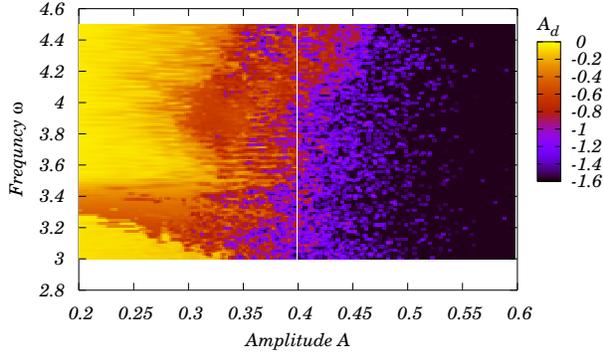}
\caption{Minima of $A_d$ vs frequency $\omega$ and amplitude $A$ of radiation in case of a standing wave.
 \label{f6}}
\end{figure}

Although the case described above has a nice physical interpretation, because we switch on the light in one particular moment, it is very difficult  to explain it in details using a simplified method presented in the next section.
The reason is that the simplified equations for $A_d$ are derived for monochromatic wave coming from both sides.
When we consider the boundary conditions described in the beginning of the present section we find that the radiation has a form of a very wide wavelet. 
Despite nonlinearities  in the equation the wavelet is always a superposition of many monochromatic waves.
Although we can assume that after a long time we have almost 
a single frequency standing wave there is still a problem 
of initial conditions for $A_d(t_0)$. 
We found that the evolution of $A_d(t)$ highly depends on the initial conditions. 
We cannot tell whether $A_d$ would oscillate or jump to $-\frac{\pi}{2}$ (and force the creation of soliton pairs).\\
It is much easier to consider the Cauchy problem for conditions:
$\xi(t=0,x)=Ar_k(t=0,x)$, where $r_k(t,x)$ is a real combination of scattering modes $\eta_{\pm k}$ in a form of a standing wave $r_k(t,x)=h_k(x)\cos\omega t$ where:
\begin{equation}
h_k(x)=\frac{\left(3\tanh^2x-1-k^2\right)\sin kx-3\tanh x\cos kx}{\sqrt{\left(k^2+4\right)(k^2+1)}}.
\end{equation}
The denominator was introduced just for convenience so that the amplitude of the wave far away from the kink would be 1.
Basically the structure of the solutions look the same but for different parameters $A$ and the oscillations begin for $t=0$.
We have plotted an analogical Figure to \ref{f5} (Figure \ref{f6}). We can see that the border between the two types of solution (with a creation (dark spots) and without) also seems to be fractal.

%The result is plotted in Figure \ref{f5} (the first case) and Figure \ref{f6} (the second case).
If we make a similar figure using the longer evolution the border would be shifted in the direction of smaller $A$ for obvious reasons. 
We tried to evolve our system as long as the border does not change significantly with longer time. 
Unfortunately this procedure was very difficult to apply and we cannot be certain if the border presented in Figure \ref{f6} is a true limit, but we think it is a quite good approximation nevertheless.

\subsection{Theory}

Let us consider a simplified field equation for $\xi$ in the $\phi^4$ 
model:
\be
\ddot{\xi}+\opL\xi+6\phi_s\xi^2=0. \label{eq11}
\ee
We have neglected only the cubic term.\\
Let $\xi$ has a form
\be
\xi(t,x) = Ar_k(t,x)+A_d(t)\eta_d(x)+\eta(t,x),\label{eq11_a} 
\ee
where $\eta$ is orthogonal to $\eta_d$.
It is consistent to assume that $A_d$ and $\eta$ are both of order
${\cal O}(A^2)$.
After substitution (\ref{eq11_a}) into (\ref{eq11}) we obtain (in the second order in $A$):
%The whole equation for $\eta_d$ yields
%In this order the equation has a form:
\begin{equation}
%\begin{eqnarray}
\begin{array}{l}
%\lefteqn{
\left(\ddot{A_d}+\omega_d^2A_d\right)\eta_d+\ddot{\eta}+\opL\eta+\\
+6\phi_s \biggl(\frac{1}{2}A^2h_k^2\left(1+\cos2\omega t\right)
    +A_d^2\eta_d^2+\eta^2+\\
+2A_d\eta_dAh_k\cos\omega t
+2A_d\eta\eta_d+2Ah_k\eta\cos\omega t\biggr)=0.
%\end{eqnarray}
\end{array}
\label{eq12_1}
\end{equation}

The first correction to $A_d$ originated from $\eta$ would be of order
${\cal O}(A_d^3)$ and now we will neglect this term.
Since $\eta$ is orthogonal to $\eta_d$ we can multiply both sides of
this equation by $\eta_d$ and integrate (similar procedure was applied in \cite{Kiselev}). The projected equation
yields:
\be
  \ddot{A_d}+\omega_d^2A_d+A^2\alpha(k)\left(1+\cos2\omega
t\right)+\beta(k)AA_d\cos\omega t+\gamma A_d^2=0, \label{eq12}
\ee
where
\be
\alpha(k)=3\frac{\displaystyle\int_{-\infty}^\infty\!\!\!\!dx\;\phi_s(x)h^2_k(x)\eta_d(x)}
 {\displaystyle\int_{-\infty}^\infty\!\!\!\!dx\;\eta_d^2(x)},
\ee
\be
\beta(k)=12\frac{\displaystyle\int_{-\infty}^\infty\!\!\!\!dx\;\phi_s(x)h_k(x)\eta_d^2(x)}
 {\displaystyle\int_{-\infty}^\infty\!\!\!\!dx\;\eta_d^2(x)},
\ee
\be
\gamma=6\frac{\displaystyle\int_{-\infty}^\infty\!\!\!\!dx\;\phi_s(x)\eta_d^3(x)}
 {\displaystyle\int_{-\infty}^\infty\!\!\!\!dx\;\eta_d^2(x)} =
\frac{9\pi}{16}.
\ee

The first two integrals can also be calculated
analytically:
\be
  \alpha(k)=\frac{9\pi}{64N^2}\left(8k^4+34k^2+17\right)
\left(1-\frac{1}{\cosh k\pi}\right)
%\frac{18}{32N^2}\pi(8k^4+24k^2+17)-18\pi\frac{32k^6-104k^4+52k^2+17}{32N^2\cosh k\pi}
\ee
\be
  \beta(k)=3\pi k^2\frac{k^4+2k^2-8}{8N\sinh\frac{k\pi}{2}}
%18\pik\frac{11k^5-6k^4-110k^3+120k^2+104k-144}{60N\sinh\frac{k\pi}{2}},
\ee
where $N=\sqrt{(k^2+4)(k^2+1)}$.\\

%The above equation is a simple forced  harmonic oscillator equation
%and we can solve it very easy:
Equation  (\ref{eq12}) is very similar to Mathieu's equation but with extra driving force and nonlinear term.
This is just a simplification because
the oscillating mode couples to the scattering modes and is a source of radiation itself(\cite{Manton}, \cite{trom}). 
This radiation carries away the energy from the discrete mode causing the decay of the mode and that means we should include a damping term in (\ref{eq12}). 
We could predict the loss of this energy in the similar way as we did in \cite{trom} but in this case there are some problems. 
The nonlinear part in eq. (\ref{eq12_1}) contains sources in the form of $r_kA_d\eta_d$ which means that we do not know how much energy is taken from the oscillational mode and how much from original radiation. 
Finally we do not know how this loss would decrease amplitudes for different frequencies of the oscillational mode. 
This is why we did not add the damping term. 
%However we hope we can find some critical conditions for the creation process.

%There is no damping term so we do not expect decaying solutions as shown in Fig. \ref{f2} but we hope we can find some critical conditions for creation process.\\
Let us find a solution of the equation in a form (there are no free oscillations of the discrete mode $\eta_d$ i.e. there are no ${\cal O}A^0$ and ${\cal O}A^1$ terms):
$$A_d=A_d^{(2)}A^2+A_d^{(3)}A^3+A_d^{(4)}A^4+\cdots$$
with a condition that $A$ is very small.
The first equation is of order ${\cal O}(A^2)$:

\be
\ddot{A}_d^{(2)}+\omega_d^2A_d^{(2)}+\alpha (1+\cos2\omega t)=0.
\ee
This equation is a simple forced harmonic oscillator equation and we
can solve it very easily:

\be
A_d^{(2)}(t) =
-\frac{\alpha}{\omega_d^2}+\frac{\alpha}{4\omega^2-\omega_d^2}\cos2\omega t+B^{(2)}\cos(\omega_dt+\delta).
\ee
We can also apply initial conditions such as $A_d(0)=0,
\;\dot{A_d}(0)=0$ and from that we can calculate $B^{(2)}$ and
$\delta$. Let us assume $\delta=0$.
We can write the next order equation:
\be
 \ddot{A}_d^{(3)}+\omega_d^2A_d^{(3)}+\beta A_d^{(2)}\cos\omega t=0,
\ee
 and solve it as well.
The inhomogeneous term is equal to:
%begin{}
\begin{equation}
%{eqnarray}
\begin{array}{r}
 \displaystyle \beta A_d^{(2)}\cos\omega
t=\alpha\beta\left(-\frac{\cos\omega t}{\omega_d^2}+
  \frac{\cos3\omega t+\cos \omega
t}{2\left(4\omega^2-\omega_d^2\right)}
  \right)+\\
\\
+\frac{1}{2}B^{(2)}\beta\bigl
(\cos(\omega_d+\omega)t+\cos(\omega-\omega_d)t\bigl).
%\end{eqnarray}
\end{array}
\end{equation}

This is a source for harmonic equation. The solution would oscillate
with the following frequencies: $\omega, 3\omega, \omega_d+\omega,
\omega-\omega_d$ (compare with Figure \ref{f4}). Of course there is a special case when
$\omega=2\omega_d$. We have then the source with resonance frequency
and the linear growth of the amplitude and therefore the small
amplitude approximation fails. In fact this approximation also fails
if $\omega^2-4\omega_d^2$ is very small, because in the solution
there is a term $\frac{1}{\omega^2-4\omega_d^2}$ and although $A$ is
small $A_d$ is no longer small.\\
The most interesting question now is how big the amplitude $A_d$ can in
be.
We neglected some terms which are important to solve this problem.
First of all this is only the solution in ${\cal O}(A^3)$ order.
Numerical solutions of the equation (\ref{eq12}) show that the most important role
plays the quadratic term in $A_d$.
We also neglected the cubic term and of course the field $\eta$ which
is the field of radiation escaping from this oscillating mode.
We do not know how small $A$ should be around the resonance frequency
so that the oscillating mode amplitude would remain small during the whole
evolution.

\begin{figure}
\PutFigure{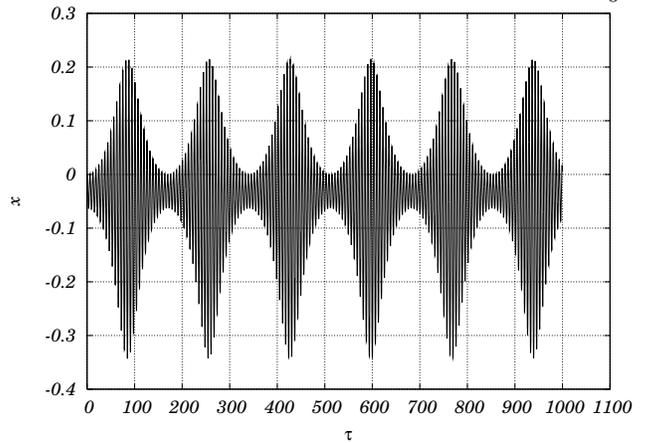}
\caption{Solution of the eq. \rev{OEQ1} for $g_1=0.2,\,g_2=0.03$. Low
frequency modulation  with large amplitude.
\label{f7}
\vspace*{5mm}}
\end{figure}

Let us now focus on the equation for $\omega=2\omega_d$.
Although we can find the precise values of the coefficients
$\alpha(k)$
and $\beta(k)$ for $k=2\sqrt{2}$ for this particular model we think it
is better to examine the whole equation for any set of coefficients.
For different models there should be different values but the 
structure of the equation remains the same.
First we simplify the equation in order to get rid of too many
parameters.
We introduce
\be
	\left\{\begin{array}{ll}
		w =&  \frac{\omega_d^2}{\gamma}A_d\\
		\tau =& \omega_d t\\
		g_1 = & \frac{\beta}{\omega_d^2}A\\
		g_2 = & \frac{\alpha\gamma}{\omega_d^4}A^2
	\end{array}\right.
\ee
and our equation takes a form:
\be
	\ddot{w}+w+w^2+g_1w\cos2\tau+g_2(1+\cos4\tau)=0,\label{OEQ1}
\ee
where $\dot{}$ denotes $\frac{d}{d\tau}$
Solutions of the above equation can be divided into three groups. In
the first group the solution oscillates around some $x_0$ with
frequencies $\omega, 2\omega,\cdots$ and is modulated with a very
small frequency. The solutions have sometimes quite large amplitude,
and the maximum is well above 0 (Figure \ref{f7}).

\begin{figure}
\PutFigure{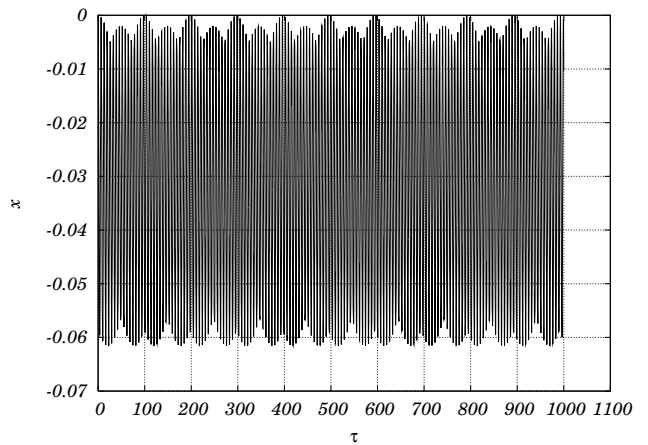}
\caption{  Solution of the eq. \rev{OEQ1} for $g_1=0.1,\,g_2=0.03$.
High frequency modulation with small amplitude.
\label{f8}
\vspace*{5mm}}
\end{figure}

The second group are solutions that oscillate between some minimum and
0. The modulation frequency can be both low and high (Figure
\ref{f8}).

\begin{figure}
\PutFigure{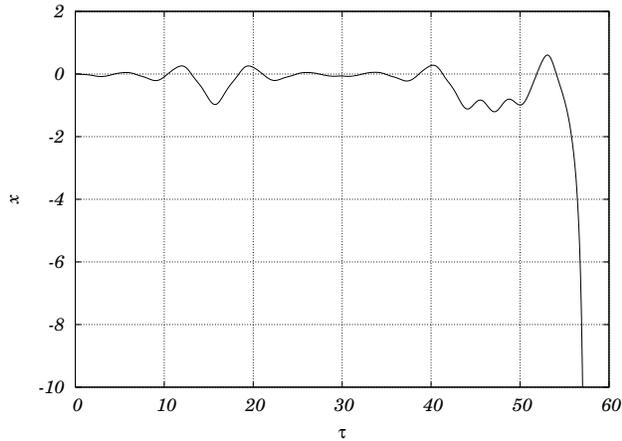}
\caption{Solution of the eq. \rev{OEQ1} for $g_1=0.9,\,g_2=0.03$.
Singularity around $t=57$.
\label{f9}
\vspace*{5mm}}
\end{figure}

The third group of these solutions are solutions which explode in a
certain time $\tau_{cr}$, they are singular (Figure \ref{f9}).

For large modulus of $w$ we can neglect linear terms in $w$ and focus
on the equation:
\be
	\ddot{w}+w^2=0,
\ee
we can find the behavior of the solution for very large $|w|$:
$$
w\sim-6(\tau-\tau_o)^{-2}
$$
Of course we cannot extrapolate these solution exactly to our partial equation for obvious
reasons. 
First of all the amplitude in a field theory cannot grow
unlimited, but in this case pairs of soliton and antisoliton are
created. 
This process can also happen when the amplitude is large
enough.
%\begin{center}
%{\Large \it Tutaj tez cos nie gra w tym akapicie ponizej}
%\end{center}

\begin{figure}
\PutFigure{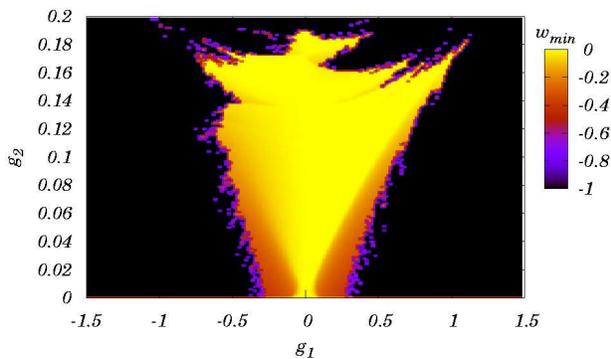}
\caption{Stable and unstable (dark) solutions on a plain 
$(g_1,g_2)$ with initial conditions $w(0)=0$ and $\dot{w}(0)=0$.
\label{f10}
\vspace*{5mm}
}
\end{figure}

\begin{figure}
\PutFigure{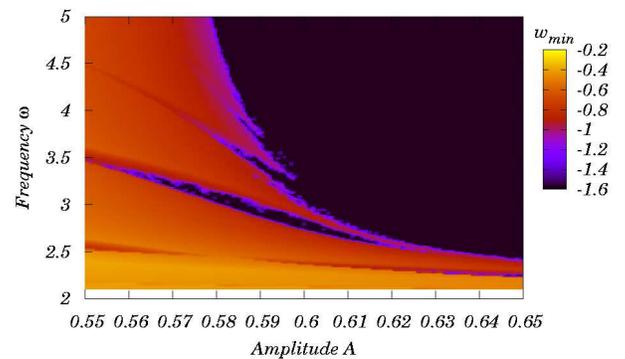}
\caption{Minima of the solutions of the simplified equation in a plain of amplitude and frequency.
\label{f11}
\vspace*{5mm}}
\end{figure}

There is also one more interesting observation according this 
simplified equation. 
The boundary between regular and singular solutions 
on a plain $g_1,\,g_2$ has a very complicated, presumably fractal, shape (Figure \ref{f10}). The same feature possesses the border between regular and singular solutions on a phase space of initial conditions with fixed $g_1$ and $g_2$ (Figure \ref{f11}).\\
In fact it is not surprising. When we take a closer look at the equation  (\ref{eq12}) we can see it possesses an invariance under discrete translation in time $t\rightarrow t+T$, where $T=\frac{2\pi}{\omega}$. If we knew the map in phase space $F:(A_d(t),\dot{A}_d(t))\rightarrow(A_d(t+T),\dot{A}_d(t+T))$ we could just iterate $F^n$ and find out whether for $n\rightarrow\infty$ the solutions are singular or not. The same procedure one usually applies when one wants to obtain a Julia or Mandelbrot sets some of which are one of the most known fractals.
\begin{figure}
\PutFigure{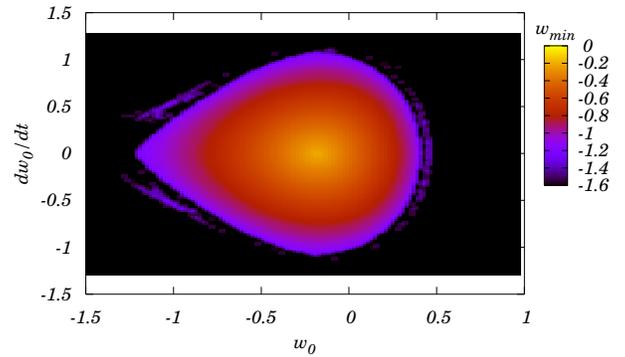}
\caption{Stable and unstable solutions on a plain 
$(w_0,\dot{w}_0)$ for fixed $g_1=0.5$ and $g_2=0.12$.
\label{f10_1}
\vspace*{5mm}}
\end{figure}
%As a final result we present the Figure \ref{f11} which shows the boundary between solutions of the equation (\ref{eq12}) which oscillate with an amplitude smaller than 
%$\frac{\pi}{2}$ and those solutions which cross the value $A_d=-\frac{\pi}{2}$. In the second case a pair of kink antikink is expected to be produced. 
As a final result we present the Figure \ref{f11} which shows the minima of the solutions of the simplified equation (\ref{eq12}). Dark regions as usual represent the unstable solutions which certainly cross the level of $-\frac{\pi}{2}$.
For those conditions we expect a pair of kink and antikink is produced.

After comparing these results with the results obtained for full partial differential equation (Fig \ref{f5} and \ref{f6}) we can see that there are of course discrepancies but the theoretical boundary lies not very far away from the true boundaries. 
It is actually easy to explain if we consider the energy which is radiated from the oscillational mode. In the theoretical picture we neglected this radiation.
%We should add some extra damping term to our simplified equation but we do not know exactly how to do this.
Of course we also did not include the cubic term and we neglected the orthogonal modes, but even in that oversimplified picture it is clear that something interesting happens.

There is also one more  interesting thing. Both in \cite{Matzner} and \cite{Weinstein} the simplified theories which were based upon collective coordinates reproduced the fractal structure which was wider than in field theory. 
In our case we used the projection onto the internal mode of the soliton and our approach gave the structure which was much more narrow than in the partial differential equation.
%Usually one can expect that the field theory would smooth the shapes but in our case it is the opposite.

\section{Conclusions and discussion}

In the present paper we have presented the evolution of a kink with an external perturbation (radiation) in $\phi^4$ theory. 
We found that during the evolution an oscillating mode is being excited.
We measured the excitation using numerical methods for two sets of initial and boundary conditions.
Depending on the frequency of the radiation and its amplitude the excitation of the oscillational mode can remain small enough to be treated only as a deformation of the kink or can grow and finally can effect in a production of kink and antikink which are radiated into the spatial infinity.
We tried to explain this process using a simplified method by projecting our field theory equation onto the oscillational mode. 
T he obtained equation is a nonlinear ordinary equation which we cannot solve analytically, but we may study it numerically. 
This equation possesses very rich space of solutions. 
Some of them are regular ones and some of them are singular. 
The border between those types of solutions is a very complicated both in a space of parameters and initial conditions. 
The solutions of the simplified equation are relatively good approximations of the solutions of the whole partial equation.
The most interesting case of creation of kink-antikink pair reveals also a quite well agreement between field theory and our simplified theory.\\
There are still many unanswered questions. First  of them is why the border in the field theory is so complicated.  We do not know how to add a damping term which we think is very important to solve the previous problem.
Our simplified theory does not give the answers about the dynamics of the created defects. We do not know how much energy from the oscillational mode is converted into the kinetic energy of kinks, how much into radiation and how much energy remains in oscillating modes around all the defects.

\section{Acknowledgments}
Work performed under auspices of ESF Programme "Coslab". 
I Thank H. Arod\'z for suggesting some ideas and many hours of fruitful discussions.

%\input{p3_1}
%\newpage
%\include{bib}
%\newpage
%\include{fig}


\begin{thebibliography}{14}
\bibitem{Coslab1} Y.M. Bunkov , H. Godfin (Eds.), ``Topological Defects and the Non-Equilibrium Dynamics of Symmtry Breaking Phase Transitions'', Kluwer Academic Publ., Dordrecht/Boston/London, 2000.
\bibitem{Coslab2} H. Arod\'z , J. Dziarmaga, W.H. \.Zurek (Eds.), ``Patterns of Symmtry Breaking'', Kluwer Academic Publ., Dordrecht/Boston/London, 2003.
\bibitem{QCD} See eg. M.Baker, J.S. Ball, F. Zachariasen Phys.Rep.
{\bf 209}, 73 (1991).
\bibitem{Cosmic} T.W.B. Kibble, J.Phys. {\bf A9}, 1387 (1979).
\bibitem{Saka} F.R. Bouchet, P. Peter, A. Riazuelo, M. Sakellariadou,
 Phys.Rev. {\bf D65}, 021301 (2002).
\bibitem{Kiselev} V.G. Kiselev, Ya,M. Shnir Phys. Rev. {\bf D57}, 5174
(1998).
\bibitem{trom}T. Roma\'nczukiewicz, Acta Phys. Pol. {\bf  E35},
523 (2004).
\bibitem{Matzner}P. Anninos, S. Oliveira, R.A. Matzner, Phys. Rev. {\bf D44} 1148 (1991).
\bibitem{Gleiser} M. Gleiser, Phys. Rev. {\bf D49}, 2978 (1994).
\bibitem{Weinstein}R.H. Goodman, P.J. Holmes, M.I. Weinstein, Physica {\bf D161}, 21 (2002).
\bibitem{Manton} N.S. Manton, H. Merabet, Nonlinearity {\bf 12}, 851
(1997).
\bibitem{Arodz} H. Arod\'z  Acta Phys. Pol. {\bf  B33}, 1241 (2002).
\bibitem{Pelka} R. Pe\l ka  Acta Phys. Pol. {\bf  B28}, 1981 (1997).
\bibitem{Slusarczyk} M. \'Slusarczyk  Acta Phys. Pol. {\bf  B31}, 617 (2000).
\end{thebibliography}
\end{document}